\documentclass[aps,prl,twocolumn,groupedaddress,superscriptaddress]{revtex4-2}
\usepackage{graphicx}
\usepackage{amssymb,amsmath}
\usepackage{epstopdf}
\usepackage{float}
\usepackage{xcolor}
\usepackage{pifont}
\usepackage{bm}
\usepackage{mathrsfs}
\usepackage{bbold}
\usepackage{upgreek}
\usepackage{physics} 
\usepackage{array}
\usepackage{times}
\usepackage{url}
\usepackage{xcolor}

\begin{document}

\title{Accelerating evaporative cooling of a strongly interacting Fermi gas by tilting the optical trap with a magnetic field gradient}

\author{Bolong Jiao}
\thanks{These authors contributed equally to this work.}
\affiliation{School of Physics and Astronomy, Sun Yat-sen University, Zhuhai, Guangdong, China 519082}

\author{Shuai Peng}
\thanks{These authors contributed equally to this work.}
\affiliation{School of Physics and Astronomy, Sun Yat-sen University, Zhuhai, Guangdong, China 519082}

\author{Qinxuan Peng}
\thanks{These authors contributed equally to this work.}
\affiliation{School of Physics and Astronomy, Sun Yat-sen University, Zhuhai, Guangdong, China 519082}

\author{Shaokun Liu}
\affiliation{School of Physics and Astronomy, Sun Yat-sen University, Zhuhai, Guangdong, China 519082}

\author{Mengde Gan}
\affiliation{School of Physics and Astronomy, Sun Yat-sen University, Zhuhai, Guangdong, China 519082}

\author{Jiaming Li}
\email[]{lijiam29@mail.sysu.edu.cn}
\affiliation{School of Physics and Astronomy, Sun Yat-sen University, Zhuhai, Guangdong, China 519082}
\affiliation{Guangdong Provincial Key Laboratory of Quantum Metrology and Sensing, Sun Yat-Sen University, Zhuhai 519082, China}
\affiliation{Shenzhen Research Institute of Sun Yat-Sen University, Shenzhen 518057, China}
\affiliation{State Key Laboratory of Optoelectronic Materials and Technologies, Sun Yat-Sen University, Guangzhou 510275, China}

\author{Le Luo}
\email[]{luole5@mail.sysu.edu.cn}
\affiliation{School of Physics and Astronomy, Sun Yat-sen University, Zhuhai, Guangdong, China 519082}
\affiliation{Guangdong Provincial Key Laboratory of Quantum Metrology and Sensing, Sun Yat-Sen University, Zhuhai 519082, China}
\affiliation{Shenzhen Research Institute of Sun Yat-Sen University, Shenzhen 518057, China}
\affiliation{State Key Laboratory of Optoelectronic Materials and Technologies, Sun Yat-Sen University, Guangzhou 510275, China}

\date{\today}

\begin{abstract}

We present a rapid evaporative cooling scheme for a strongly interacting $^{6}\mathrm{Li}$ Fermi gas in an optical dipole trap. The method uses a magnetic-field-gradient--induced tilt of the trapping potential to accelerate cooling in the unitarity-limited regime. In evaporation based only on lowering the optical trap depth, the unitarity-limited scattering cross section can support runaway cooling; however, the cooling rate slows around $T/T_F \simeq 0.5$, and the runaway behavior is no longer maintained. We improve on this approach by applying a magnetic-field gradient when the gas temperature reaches about half the Fermi temperature. The induced tilt opens an escape channel for energetic atoms while keeping the trap frequencies nearly unchanged. This modification increases the cooling speed and cools the gas below the superfluid transition temperature, reaching $T/T_F = 0.16$ on a timescale of $\sim 25\,\mathrm{ms}$. Our results provide a simple and robust route for rapidly cooling a strongly interacting Fermi gas into the superfluid regime, facilitating studies of the physics of unitary Fermi superfluids.

\end{abstract}

\maketitle

\section{I. Introduction}

Evaporative cooling is a fundamental technique for producing quantum-degenerate atomic gases~\cite{C. S. Adams1995, W. Ketterle1996, O. J. Luiten1996, JWeiner1999, OHara2001, K. M. OHara2002, J. F. Cl2009}. Its efficiency relies on maintaining a high elastic collision rate while selectively removing energetic atoms, and can be improved through various methods, including unitarity-limited cooling near a Feshbach resonance~\cite{L. Luo2006}, trap-geometry optimization~\cite{T. Kinoshita2005, K. J. Arnold2011, M.-S. Heo2011, Abraham J. Olson2013, M. J. Williams2015, I. Fritsche2021}, tilting the trap with a magnetic gradient~\cite{C.-L. Hung2008, J. Zeiher2021}, and three-body cooling~\cite{Peng2024}. 
The demand for faster and more efficient cooling has also intensified with ultracold molecular gases~\cite{H. Son2020, M. Duda2023},  systems with strong inelastic collisions ~\cite{peng2023,S. Peng2025, LiuArxiv2025}, and atoms in box-shaped potentials~\cite{A. L. Gaunt2013, N. Navon2021}. These systems therefore benefit from protocols that enable rapid evaporation while sustaining high elastic collision rates, particularly when the available lifetime is limited. To overcome this difficulty, one may adopt a hybrid cooling scheme.

Motivated by earlier studies of unitarity-limited cooling~\cite{L. Luo2006} and runaway tilt cooling in weakly interacting Bose gases~\cite{C.-L. Hung2008}, we present a rapid evaporative cooling scheme for a strongly interacting $^{6}\mathrm{Li}$ Fermi gas in an optical dipole trap using a hybrid cooling method, in which a magnetic-field-gradient-induced tilt of the trapping potential is applied to accelerate cooling in the unitarity-limited regime. In evaporative cooling based only on lowering the optical trap depth, the unitarity-limited scattering cross section can support runaway cooling; however, the cooling rate slows around $T/T_F \approx 0.5$~\cite{L. Luo2006}, and the runaway evaporation is no longer maintained. 
We improve on this approach by applying a magnetic-field gradient when the gas temperature reaches about half the Fermi temperature. The resulting magnetic tilt opens a directional escape channel, while the strong interactions near the broad $s$-wave Feshbach resonance maintain near-maximal collision rates. Thus, this method preserves the high trap frequencies of the optical dipole trap, allowing rapid rethermalization.
As a result, this hybrid scheme increases the cooling speed and cools the gas below the superfluid transition temperature, reaching $T/T_F = 0.16$ on a timescale of $\sim 25\,\mathrm{ms}$. The scheme also reduces the influence of the large bias magnetic field required to stay at unitarity on very shallow optical traps. 

Our results provide a simple and robust route for rapidly cooling a strongly interacting Fermi gas into the superfluid regime, facilitating future studies of collective flow, shock wave dynamics, and other far-from-equilibrium phenomena in strongly interacting systems~\cite{J. A. Joseph2011, J. Eggers2008}.

\section{II. EXPERIMENTS and RESULTS}

\begin{figure}[htbp]
	\begin{center}
		\includegraphics[width=0.9\columnwidth, angle=0]{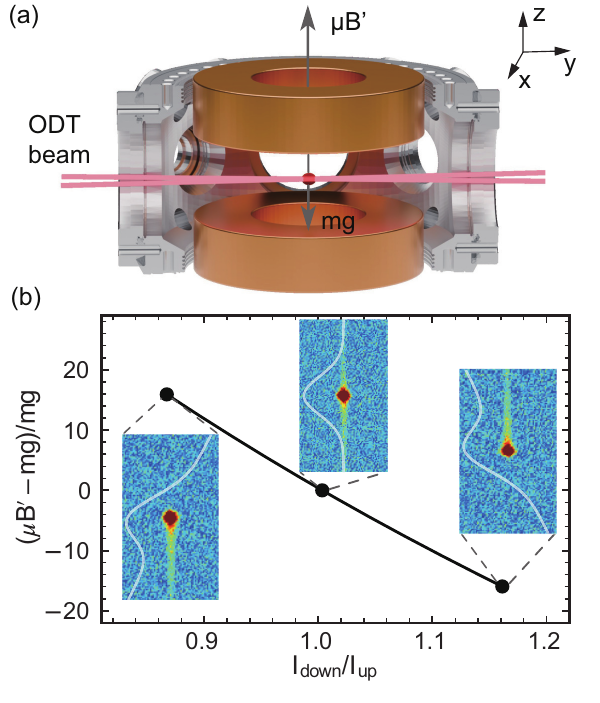}
		\caption{Schematic of the system setup.
			(a) Diagram of the crossed-beam ODT, magnetic coils, and vacuum chamber. 
			(b) Illustration of the MGT that tilts the total potential along the vertical $z$ axis. The in-situ column density of the atomic cloud is monitored for different values of the magnetic driving current ratio $I_{\mathrm{down}}/I_{\mathrm{up}}$, which controls both the strength and the direction of the magnetic tilt. The white curve schematically indicates the trapping potential. 		
		} \label{fig:experimental_setup}
	\end{center}
\end{figure}

We begin by describing the implementation and basic properties of the magnetic-gradient trap (MGT) used to tilt the optical potential.
Fig.~\ref{fig:experimental_setup}(a) shows the layout of the apparatus for ultracold $^{6}$Li, including the crossed optical dipole trap (ODT) and a pair of high-stability Helmholtz coils that generate a homogeneous bias magnetic field along the vertical $z$ axis~\cite{X. Zhang2019, ghong2021, Q. Peng2025, H. Liu2023}. A controlled imbalance in the currents through the two coils produces a tunable magnetic-field gradient $B'$, which is superimposed on the bias field and gives rise to the MGT potential $U_\text{grd}$. In Appendix~A, we present a detailed theoretical model of this magnetically tilted potential and derive the corresponding trap-frequency scaling. This analysis shows how the tilt shifts the minimum of the trapping potential while maintaining high trapping frequencies. 

Both high-field-seeking hyperfine states $\ket{1}$ and $\ket{2}$ of $^{6}$Li ~\cite{Li2016} experience a repulsive potential in the relevant field range, so the applied gradient generates a positive tilt $U_\text{grd}>0$ along the $z$-axis. By varying the current ratio $I_{\mathrm{down}}/I_{\mathrm{up}}$, the direction and magnitude of the tilt can be precisely controlled, enabling the creation of a tunable ``magnetic bowl'' potential. The tilt opens an escape channel for energetic atoms in the downhill direction and allows real-time observation of the escaping flux, as illustrated in Fig.~\ref{fig:experimental_setup}(b).

The smallest adjustable step in $I_{\mathrm{down}}/I_{\mathrm{up}}$ corresponds to a resolution of 0.04~$\mathrm{\mu}$K in effective trap depth. With a maximum current of 140~A, we achieve the range $0.867 \le I_{\mathrm{down}}/I_{\mathrm{up}} \le 1.161$, corresponding to the largest tilt used in this work. Appendix~B provides details of the coil model, current calibration, and the reconstruction of $B'(t)$ and $U(t)$. At maximum tilt, the effective trap depth along the downhill direction is reduced to 1.2~$\mathrm{\mu}$K, while the nominal optical trap depth is held at $U_\text{opt} = 6.6~\mathrm{\mu}$K near the broad Feshbach resonance at 832~G. Under these conditions, the maximum allowed force ratio $(\mu B' - m g)/(m g) = 16$, enabling highly directional evaporation. Here, $\mu = \mu_\text{B}$ is the Bohr magneton.

\begin{figure}[htbp]
	\begin{center}
		\includegraphics[width=0.9\columnwidth, angle=0]{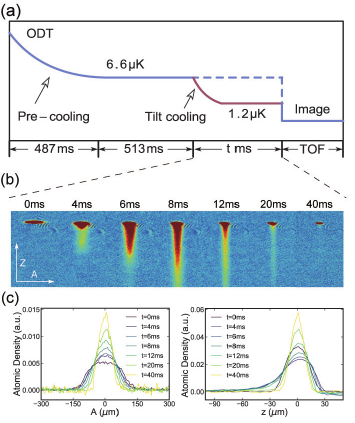} 
		\caption{Experimental sequence and evolution.
			(a) Timing sequence for magnetic tilt-assisted evaporative cooling. 
			The blue curve shows the ODT-depth ramp during the pre-cooling stage, and the red curve marks the subsequent cooling phase with the MGT activated.
			(b) \textit{In situ} absorption images taken after applying the MGT, demonstrating the evolution of the atomic cloud and the directional escape of energetic atoms along the tilt axis.
			(c) Corresponding one-dimensional column-density profiles along the vertical $z$ and horizontal $A$ directions.		
		}\label{fig:experimental_procedure}
	\end{center}
\end{figure}

Fig.~\ref{fig:experimental_procedure}(a) shows the experimental timing sequence. After laser cooling, atoms are loaded into the ODT and the magnetic field is rapidly ramped to 832~G, preparing a balanced mixture of $\ket{1}$ and $\ket{2}$. The ODT is then lowered from 380~$\mathrm{\mu}$K to 6.6~$\mathrm{\mu}$K over 1~s, yielding $T = 0.4~\mathrm{\mu}$K, $N = 5.5 \times 10^5$ (per spin state), and   $T/T_F = 0.28$. The measured trap frequencies are
$2\pi \times (580 \pm 20,\ 62 \pm 5,\ 580 \pm 20)$~Hz. The magnetic-tilt cooling phase follows. At $t=0$, the maximum gradient ratio of $\left(\mu B^{\prime} -mg\right) /mg = 16$ is applied and held for a variable duration $t$. Due to eddy currents (see Ref.~\cite{Y. Chen2021}), the gradient reaches its steady value on a 5-ms timescale and stabilizes within 25~ms. After the cooling stage, the atom number $N$ and temperature $T$ are extracted from absorption images taken after 3~ms of hydrodynamic expansion. A detailed description of the temperature-extraction procedure is provided in Appendix~E.

\begin{figure}[h]
	\begin{center}
	\includegraphics[width=1.0\columnwidth, angle=0]{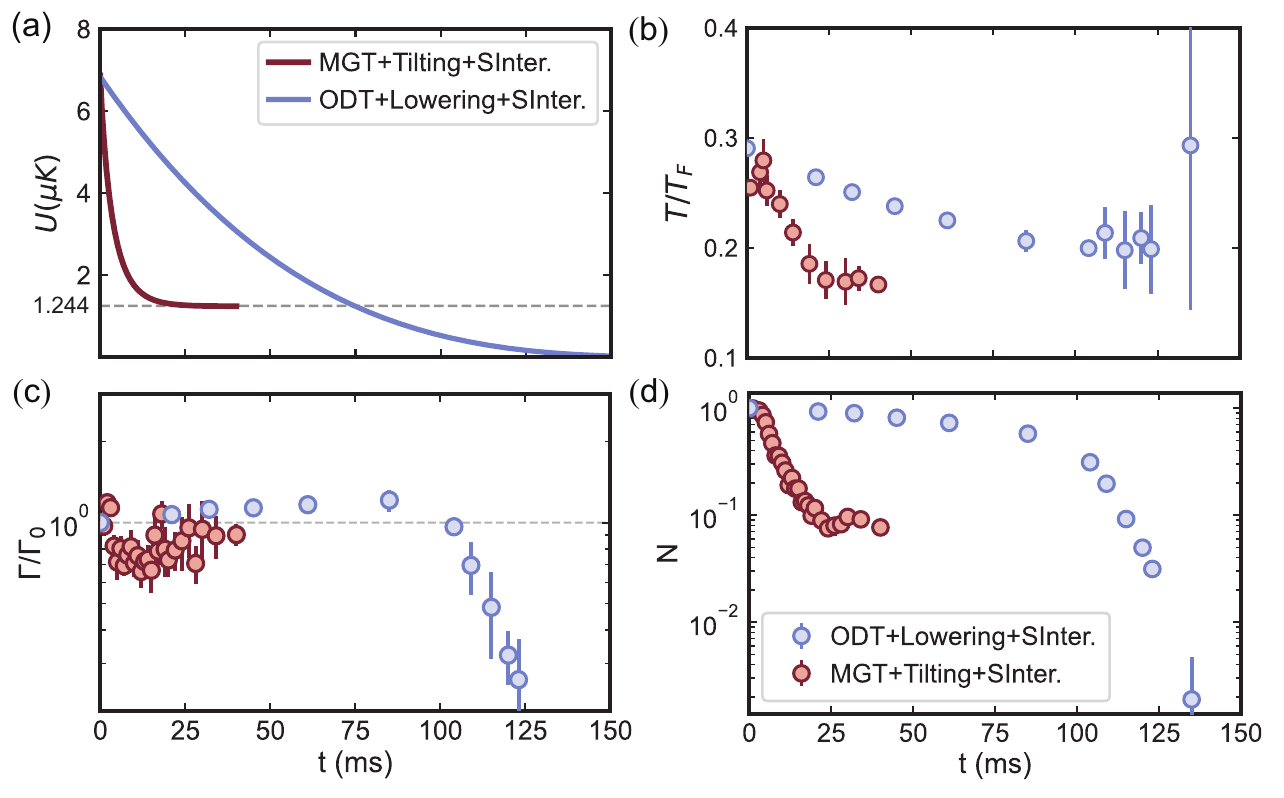}
	\caption{Comparison for the magnetically tilted cooling (red) and the conventional optical-trap-lowering cooling (blue) in the unitary regime.
		(a) Effective trap depth $U(t)$.
		(b) Reduced temperature $T/T_{F}$, showing that the tilted protocol reaches lower degeneracy on a much shorter timescale.
		(c) Normalized collision rate $\Gamma(t)/\Gamma_0$. The tilted scheme sustains a high collision rate throughout the process, whereas the other method suffers a drop beyond $t \approx 100$~ms. The collision rate $\Gamma$ is calculated from the measured $N$ and $T$ using the unitarity-limited cross-section; error bars represent the propagated uncertainty from experimental measurements.
		(d) Normalized atom number evolution $N(t)$. The tilted protocol demonstrates stable atom-number retention as the trap depth levels off, whereas the continuous lowering of the optical trap could lead to catastrophic atom loss when the repulsive magnetic force from the bias magnetic field for Feshbach resonance becomes comparable in magnitude to the optical confining force.
	}\label{fig:tilt_0.015U0_t_results}
	\end{center}
\end{figure}

Figs.~\ref{fig:experimental_procedure} (b) and (c) show the \textit{in situ} images and corresponding density profiles, illustrating the directional atom loss and the progressive reduction of the cloud size during evaporation. The data clearly demonstrate that atoms escape through the channel opened by the magnetic tilt, leading to a rapid decrease in cloud size. These observations are in good agreement with the predictions of our model.

Fig.~\ref{fig:tilt_0.015U0_t_results} summarizes the comparison between magnetic-tilt evaporation and conventional optical-trap-lowering (ODT-lowering) in the unitary regime. The effective trap depth $U(t)$ for the tilted protocol follows an exponential decay governed by the eddy-current-limited coil response of 5~ms~\cite{Y. Chen2021}. In contrast, the ODT-lowering trajectory follows the optimized power-law form derived for strong interactions~\cite{L. Luo2006} (Fig.~\ref{fig:tilt_0.015U0_t_results}(a)).
Note that the timescale of the ODT-lowering scheme is fundamentally limited by the decreasing trap frequencies; faster ramps would violate the adiabaticity required for efficient evaporation, leading to atom loss without cooling. The reduced temperature $T/T_F$ under magnetic tilting decreases rapidly from 0.28 to 0.16 in 25~ms (Fig.~\ref{fig:tilt_0.015U0_t_results}(b)). In contrast, the ODT-lowering method is significantly slower, reaching only $T/T_F \approx 0.20$ after 100~ms, followed by a heating trend due to the loss of efficiency.

The underlying mechanism for this enhanced performance is elucidated by the evolution of the collision rate $\Gamma$ and atom number.
Here $\Gamma$ denotes the mean elastic collision rate per atom, $\Gamma = n\sigma v_{\rm rel}$, where $n$ is the atomic density, $\sigma$ is the elastic scattering cross section, and $v_{\rm rel}$ is a typical relative velocity set by the temperature~\cite{L. Luo2006}.
Crucially, the magnetic-tilt protocol maintains the optical confinement, allowing $\Gamma$ to remain comparable to its initial value $\Gamma_0$ throughout evaporation (Fig.~\ref{fig:tilt_0.015U0_t_results}(c), red circles).

Conversely, the conventional ODT-lowering scheme encounters a fundamental limit.
As the optical potential is lowered (typically below $0.8~\mu\mathrm{K}$ in our setup), the residual magnetic gradients and gravity severely distort the total trap potential, causing the trap volume to open abruptly.
This confinement collapse leads to a precipitous drop in the collision rate (Fig.~\ref{fig:tilt_0.015U0_t_results}(c), blue circles) and triggers catastrophic atom loss (Fig.~\ref{fig:tilt_0.015U0_t_results}(d)).
Consequently, runaway evaporation ceases, and the system fails to reach lower temperatures.
These results demonstrate that magnetic-tilt evaporation overcomes the trap-deformation limits of conventional schemes, offering a faster and more stable route to deep quantum degeneracy.

\section{III. Discussion}

We have demonstrated a rapid and efficient evaporative-cooling protocol for unitary Fermi gases using a magnetically tilted trap. By combining unitarity-limited elastic scattering with a controlled magnetic-gradient-induced spillway, this method sustains high collision rates throughout evaporation and achieves strong cooling on tens-of-milliseconds timescales, reducing $T/T_F$ from 0.28 to 0.16. Compared with optimized ODT lowering, the magnetic-tilt protocol reaches $T/T_F\approx0.20$ more than 75\% faster and ultimately attains significantly deeper degeneracy.

To guide the experiment, we developed a dynamical scaling model (Appendices~C and D) describing magnetically tilted evaporation in both strongly and weakly interacting regimes. The model predicts universal relations among $N$, $\Gamma$, $U$, and phase-space density $\rho$, and clarifies in more detail why tilt-based cooling benefits from the nearly constant trap frequencies. Although technical constraints---coil inductance, eddy currents, and finite gradient strength---prevented the implementation of fully optimized trajectories, the experimentally feasible ramps already approximate the theoretical optimum and achieve substantial performance gains. Improved coil designs with programmable gradients could enable sub-100 ms cooling sequences.

In Appendix~F, we further test the magnetic-tilt cooling scheme in the weakly interacting regime, where it also exhibits clear advantages in both cooling speed and the achievable final temperature. These results highlight the potential of this method for cooling and realizing degenerate gases in higher partial-wave systems---such as $p$-wave gases---where strong inelastic losses limit lifetimes and introduce heating~\cite{LiuArxiv2025}. An ultrafast cooling protocol may help overcome these limitations and open new possibilities for studies of higher partial-wave interactions.

Moreover, the directional atom outflow observed near unitarity also provides a promising tool for studying nonequilibrium and hydrodynamic transport, such as shock-wave formation, collective streams, and quantum funneling. Magnetic-tilt traps thus offer a simple and versatile platform for exploring far-from-equilibrium dynamics in strongly interacting Fermi systems.

\section*{Acknowledgements}

This work is supported by  NSFC under Grant No.12574302 and No.12174458. J.~Li received support from Fundamental Research Funds for Sun Yat-sen University 2023lgbj0 and 24xkjc015. L. Luo received support from Shenzhen Science and Technology Program JCYJ20220818102003006.


\newpage
\appendix
\renewcommand{\thefigure}{A\arabic{figure}}

\setcounter{figure}{0}

\setcounter{table}{0} 

\renewcommand{\thetable}{A\arabic{table}} 

\makeatletter
\renewcommand{\fnum@table}{\textbf{\tablename~\thetable}}
\makeatother

\renewcommand{\theequation}{A.\arabic{equation}}
\setcounter{equation}{0}
\setcounter{table}{0}
\setcounter{section}{0}
\section*{Appendix A: Modeling the magnetically tilted trap} 

The total potential along $z$-axis combines the ODT with the MGT. 
The ODT, formed by two Gaussian laser beams with a crossing angle of $2\theta$, is described by
\begin{equation}\label{eq:Uopt}
	U_{\text{opt}}(x,y,z) = -\frac{U_0}{2} \left[ e^{ -\frac{2(r_-^2 + z^2)}{w^2} } + e^{ -\frac{2(r_+^2 + z^2)}{w^2} } \right],
\end{equation}
where $U_0$ is the trap depth at the center, $r_{\pm} = x \cos \theta \pm y \sin \theta$, and $w$ is the beam waist radius.

Gravity and the magnetic gradient contribute a linear potential\textcolor{red}{:}
\begin{equation}\label{eq:Ugrad}
	U_{\text{grad}}(z) = -(\mu B'- m g) z,
\end{equation}

\begin{figure}[htbp]
	\begin{center}
		\includegraphics[width=1.0\columnwidth, angle=0]{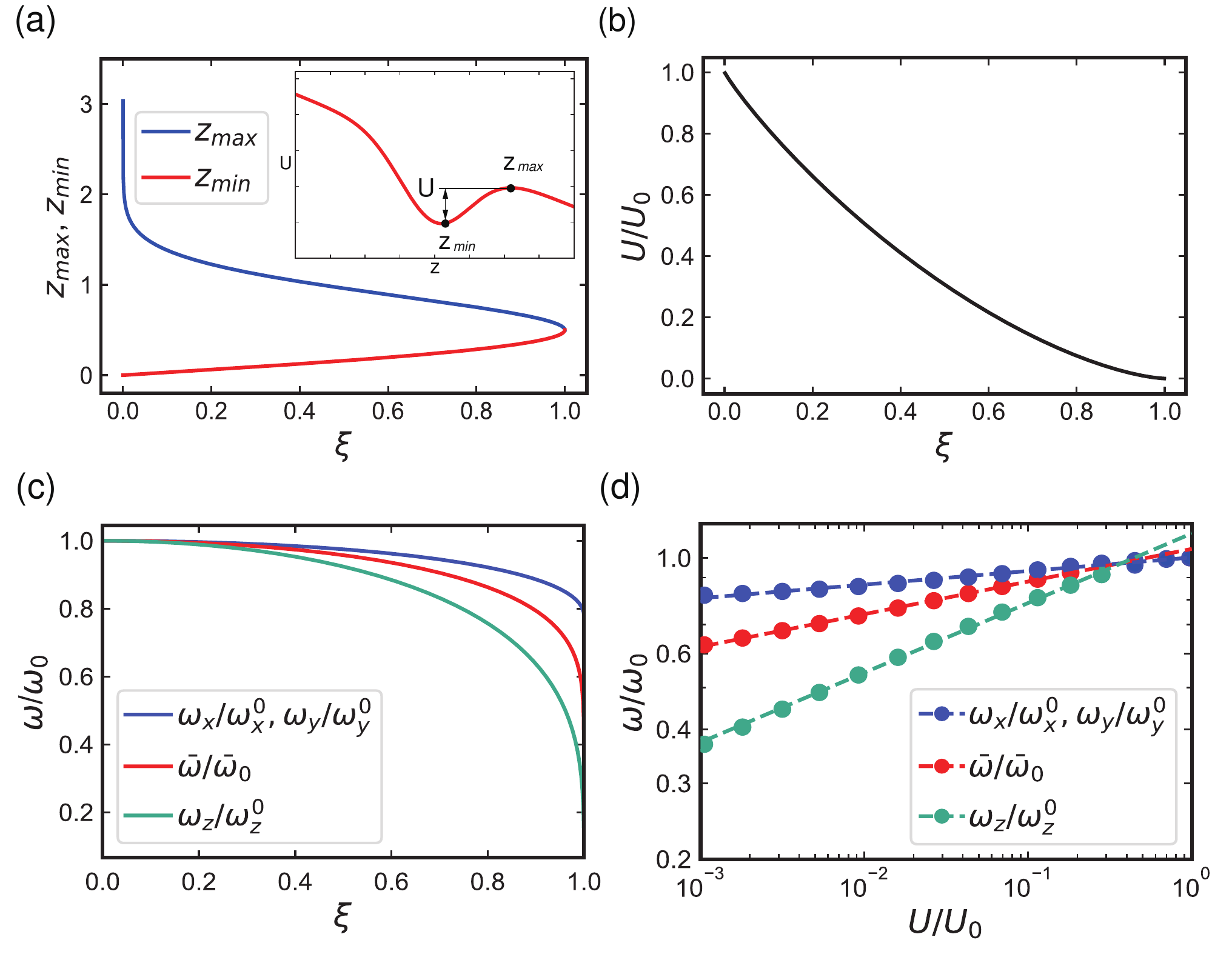}
		\caption{
			Characterization of magnetic tilted trap properties.  
			(a)-(c) show the positions of $z_{\mathrm{max}}$ and $z_{\mathrm{min}}$, the normalized trap depth $U/U_0$, and the normalized trapping frequencies $\omega/\omega_0$ as functions of the tilt parameter $\xi$, respectively.  
			(d) Scaling behavior of the trap frequencies $\omega$ with respect to the trap depth $U$. The dashed lines represent power-law fittings. Specifically, the blue line corresponds to the geometric-mean frequency ratio $\bar{\omega}/\bar{\omega}_0 \approx 1.05 (U/U_0)^{0.075}$, the red line to the transverse directions $\omega_x/\omega_x^{0} = \omega_y/\omega_y^{0} \approx (U/U_0)^{0.032}$, and the green line to the axial direction $\omega_z/\omega_z^{0} \approx 1.14 (U/U_0)^{0.162}$.
		}  \label{fig:tilt_trap_properties}
	\end{center}
\end{figure}

As shown in Fig.~\ref{fig:tilt_trap_properties}(a), the total axial potential $U_{\text{tot}}$ has a local minimum and maximum that determine the effective trap depth, defined as
\begin{equation}\label{eq:Ueff_def}
	U_{\text{eff}} = U\!\left(z_{\text{max}}\right) - U\!\left(z_{\text{min}}\right).
\end{equation}
Following Ref.~\cite{C.-L. Hung2008}, their positions are given analytically by
\begin{equation}\label{eq:zmax_zmin}
	z_\text{max} = \frac{w}{2} \sqrt{ -W_0\left( -\frac{\xi^2}{e} \right) }, \quad
	z_\text{min} = \frac{w}{2} \sqrt{ -W_{-1}\left( -\frac{\xi^2}{e} \right) },
\end{equation}
where the dimensionless tilt parameter is $\xi = \sqrt{e} \left( \mu B^{\prime} -mg \right) w / 2U_0$, and $W_0, W_{-1}$ are the two real branches of the Lambert $W$-function. When  $\xi \to 1$, the extrema merge and the trap vanishes.

Fig.~\ref{fig:tilt_trap_properties}(b) shows the normalized trap depth $U/U_0$ as a function of $\xi$. The tilted minimum at $z = z_\text{min}$ shifts the trap frequencies to
\begin{equation}\label{eq:wxwywz}
	\begin{split}
		\omega _{x,y}=\omega_{x,y}^{0}  e^{-z_\text{min}^{2}/w^2},\\
		\omega_z = \omega_{z}^{0} \sqrt{ 1 - \frac{4 z_\text{min}^2}{w^2} }  e^{-z_\text{min}^2 / w^2}.
	\end{split}
\end{equation}
where $\omega_{z}^{0}= \omega_0 = \sqrt{4U_0/(m w^2)}$, $\omega_{x}^{0} = \omega_0 \cos\theta$, and $\omega_{y}^{0} = \omega_0 \sin\theta$. The geometric-mean frequency becomes
\begin{equation}\label{eq:wxyz}
	\bar{\omega} = \bar{\omega}_0 \left( 1 - \frac{4 z_\text{min}^2}{w^2} \right)^{1/6}  e^{-z_\text{min}^2 / w^2},
\end{equation}
exhibiting only a weak dependence on the trap depth. 
A power-law fit yields
\begin{equation}\label{eq:wbar_powerlaw}
	\bar{\omega}/\bar{\omega}_0 \approx 1.05 \left( U/U_0 \right)^{0.075},
\end{equation}
highlighting a central advantage of magnetic tilting: the trap remains tightly confined even as the effective depth is greatly reduced.
Here $\bar{\omega}_0 = \omega_0 (\sin\theta \cos\theta)^{1/3}$. More details are shown in Figs.~\ref{fig:tilt_trap_properties}(c) and (d).

\section*{Appendix B: Magnetic-Field Profile and Gradient Produced by the Tilt Coils}
\label{app:tilt_coils}

We model the tilt coils as a Helmholtz-type pair of coaxial multi-turn circular coils and compute the magnetic field along the $z$-axis from the Biot--Savart law. The two coil centers are taken as
\begin{equation}\label{eq:coil_centers}
	z_{\rm d}=-\frac{h}{2},\qquad
	z_{\rm u}=+\frac{h}{2}+\delta z,
\end{equation}
where $h$ is the nominal separation and $\delta z$ is a small effective axial offset introduced to account for the as-built alignment. For an $N$-turn circular coil of radius $R$ carrying current $I$, the field along the $z$ axis is
\begin{equation}\label{eq:Bz_loop}
	B_z^{\rm (loop)}(z;z_0,I)=
	\frac{\mu_0 N I R^2}{2\bigl[R^2+(z-z_0)^2\bigr]^{3/2}},
\end{equation}
thus the total field along the $z$ axis generated by the tilt-coil pair is
\begin{equation}\label{eq:Bz_total}
	B_z(z,t)=B_z^{\rm (loop)}(z;z_{\rm u},I_{\rm up}(t))
	+B_z^{\rm (loop)}(z;z_{\rm d},I_{\rm down}(t)).
\end{equation}
The gradient along $z$ entering the linear potential term $U_{\rm grad}(z)$ in Eq.~(\ref{eq:Ugrad}) is evaluated at the instantaneous trap minimum position $z=z_c(t)$,
\begin{equation}\label{eq:Bprime_def_appB}
	B'(t)\equiv \left.\frac{\partial B_z(z,t)}{\partial z}\right|_{z=z_c(t)}.
\end{equation}
Here $z_c(t)$ corresponds to the shifted minimum position $z_{\min}$ of the tilted potential, whose analytical expression is given in Eq.~(\ref{eq:zmax_zmin}). For the dataset in Fig.~3(a), we use the effective coil parameters $R=52.820~\mathrm{mm}$, $h=129.570~\mathrm{mm}$, $\delta z=0.702~\mathrm{mm}$, and $N=107.122$.

During the tilt ramp, the tilt coils contribute to both the gradient and the local bias field experienced by the atoms. Since the trap minimum moves under tilting, the bias field is locked at the instantaneous minimum,
\begin{equation}\label{eq:B0_lock_appB_moving}
	B_z\!\bigl(z_c(t),t\bigr)=B_0,
\end{equation}
where we take $B_0=831.438~\mathrm{G}$ from RF spectroscopy (Rabi-oscillation calibration) at the cloud center. We parameterize the two currents by an upper-coil current $I_{\rm up}(t)$ and a (generally time-dependent) ratio $r(t)$ via
\begin{equation}\label{eq:Idown_ratio}
	I_{\rm down}(t)=r(t)\,I_{\rm up}(t).
\end{equation}
Defining geometry factors at the \emph{instantaneous} position $z_c(t)$ as
\begin{equation}\label{eq:beta_ud}
	\begin{aligned}
		\beta_{\rm u}(t)&\equiv \frac{\mu_0 N R^2}{2\bigl[R^2+(z_c(t)-z_{\rm u})^2\bigr]^{3/2}},\\
		\beta_{\rm d}(t)&\equiv \frac{\mu_0 N R^2}{2\bigl[R^2+(z_c(t)-z_{\rm d})^2\bigr]^{3/2}}.
	\end{aligned}
\end{equation}
the bias-lock condition Eq.~(\ref{eq:B0_lock_appB_moving}) yields
\begin{equation}\label{eq:ratio_appB_moving}
	r(t)=\frac{B_0/I_{\rm up}(t)-\beta_{\rm u}(t)}{\beta_{\rm d}(t)}.
\end{equation}
In practice, $\delta z$ is chosen such that the resulting ratio remains consistent with the current-controller outputs while maintaining the overlap between the magnetic potential center and the ODT reference position throughout the ramp.

The time dependence of $I_{\rm up}(t)$ is limited by eddy currents, and is well described by an exponential response,
\begin{equation}\label{eq:Iup_exp_appB}
	I_{\rm up}(t)=I_{\infty}-\bigl(I_{\infty}-I_{0}\bigr)e^{-t/\tau},
\end{equation}
with $\tau=5~\mathrm{ms}$ for the dataset in Fig.~3(a). In our implementation, the fitted parameters are $I_{\infty}=139.861~\mathrm{A}$ and $I_{0}=130.344~\mathrm{A}$, where $I_{\infty}$ corresponds to the maximum deliverable current of the power supply (used as the upper-limit setting during the tilt ramp). Experimentally, $I_{\rm up}(t)$ is regulated by an analog current-servo loop using a closed-loop fluxgate DC current transducer as the current feedback, yielding a calibrated transconductance $I_{\rm up}(t)=\kappa_I V(t)$ with $\kappa_I=20.208~\mathrm{A/V}$.

With $B'(t)$ obtained above Eq.~(\ref{eq:Bprime_def_appB}), the tilt parameter $\xi(t)$ is computed using its definition given in Appendix~A (immediately following Eq.~(\ref{eq:zmax_zmin})). The corresponding effective trap depth is then evaluated using Eqs.~(\ref{eq:zmax_zmin}) and (\ref{eq:Ueff_def}). This procedure yields the time-dependent $U(t)$ shown as the red curve in Fig.~3(a).

\section*{Appendix C: Comparison of the Scaling Laws}

Scaling laws provide a clear framework for comparing different evaporative-cooling schemes~\cite{O. J. Luiten1996, OHara2001}. Assuming a constant truncation parameter $\eta \equiv U/(k_B T)$, the trap depth $U$ and temperature $T$ scale proportionally. The geometric mean trap frequency follows $\bar{\omega} \propto U^{\nu}$. For conventional optical-trap-lowering  evaporation, $\nu = 1/2$ (harmonic scaling), whereas for the magnetically tilted evaporation used here, $\nu \approx 0$ \cite{C.-L. Hung2008}.

The scaling of atom number $N$, phase-space density $\rho$, and collision rate $\varGamma$ is characterized by the power laws
\begin{equation}\label{eq:scaling_laws}
	N \propto U^q, \quad \rho \propto U^d, \quad \varGamma \propto U^c.
\end{equation}
The evaporation exponent $q$ is derived from the energy balance as $q = 3(1 - \nu)/(\eta' - 3)$, where $\eta' \approx (\eta - 5)/(\eta - 4) + \eta$ in the high-$\eta$ limit~\cite{O. J. Luiten1996}. Consequently, the phase-space density exponent is $d = q + 3\nu - 3$.
The collision rate exponent $c$ depends on the interaction regime via the scattering cross section $\sigma \propto T^{-(\xi-1)}$. We define $\xi = 1$ for a weakly interacting gas (constant $\sigma$) and $\xi = 2$ for a unitary Fermi gas ($\sigma \propto T^{-1}$). Combining these with the density scaling $n \propto N \bar{\omega}^3 T^{-3/2}$ we obtain $c = q + 3\nu - \xi$.

\renewcommand{\arraystretch}{1.5}
\begin{table*}
	\centering
	\caption{\label{tab:scaling_summary}
		Summary of scaling exponents for different evaporative-cooling schemes.}
	\begin{ruledtabular}
		\begin{tabular}{lccc}
			Scheme &  $N/N_0 = (U/U_0)^q$ &  $\varGamma/\varGamma_0 = (U/U_0)^c$ &  $\rho/\rho_0 = (U/U_0)^d$ \\
			\hline
			ODT+Lowering + WInter.~\cite{OHara2001} &
			$\dfrac{3}{2(\eta' - 3)}$ &
			$\dfrac{\eta'}{2(\eta' - 3)}$ &
			$-\dfrac{3(\eta' - 4)}{2(\eta' - 3)}$ \\[2mm]
			ODT+Lowering + SInter.~\cite{L. Luo2006} &
			$\dfrac{3}{2(\eta' - 3)}$ &
			$\dfrac{6 - \eta'}{2(\eta' - 3)}$ &
			$-\dfrac{3(\eta' - 4)}{2(\eta' - 3)}$ \\[2mm]
			MGT+Tilting + WInter.~\cite{C.-L. Hung2008, J. F. Cl2009} &
			$\dfrac{3(1-\nu)}{\eta' - 3}$ &
			$\dfrac{3(1-\nu)}{\eta' - 3} + 3\nu - 1$ &
			$\dfrac{3(1-\nu)}{\eta' - 3} + 3\nu - 3$ \\[2mm]
			\textbf{MGT+Tilting + SInter. (this work)} &
			$\mathbf{\dfrac{3(1-\nu)}{\eta' - 3}}$ &
			$\mathbf{\dfrac{3(1-\nu)}{\eta' - 3} + 3\nu - 2}$ &
			$\mathbf{\dfrac{3(1-\nu)}{\eta' - 3} + 3\nu - 3}$\\
		\end{tabular}
	\end{ruledtabular}
\end{table*}

Table~\ref{tab:scaling_summary} summarizes the scaling exponents for four evaporative cooling schemes, including conventional optical trap lowering evaporation with weak~\cite{OHara2001} and strong~\cite{L. Luo2006} interactions, as well as magnetically tilted evaporation with weak interactions~\cite{C.-L. Hung2008, J. F. Cl2009} and strong interactions (this work). 
A comparison of these approaches highlights the distinct advantage of the magnetic tilt method: the near-constant trap frequencies ($\nu \approx 0$) significantly mitigate the reduction in collision rate that typically limits optical trap lowering. 
Specifically for the unitary gas, this geometry maintains a favorable collision-rate scaling $c$, enabling a continued increase in phase-space density even at shallow trap depths. 

\section*{Appendix D: TRAP-LOWERING TRAJECTORIES}

\begin{figure}[htbp]
	\begin{center}
		\includegraphics[width=1.0\columnwidth, angle=0]{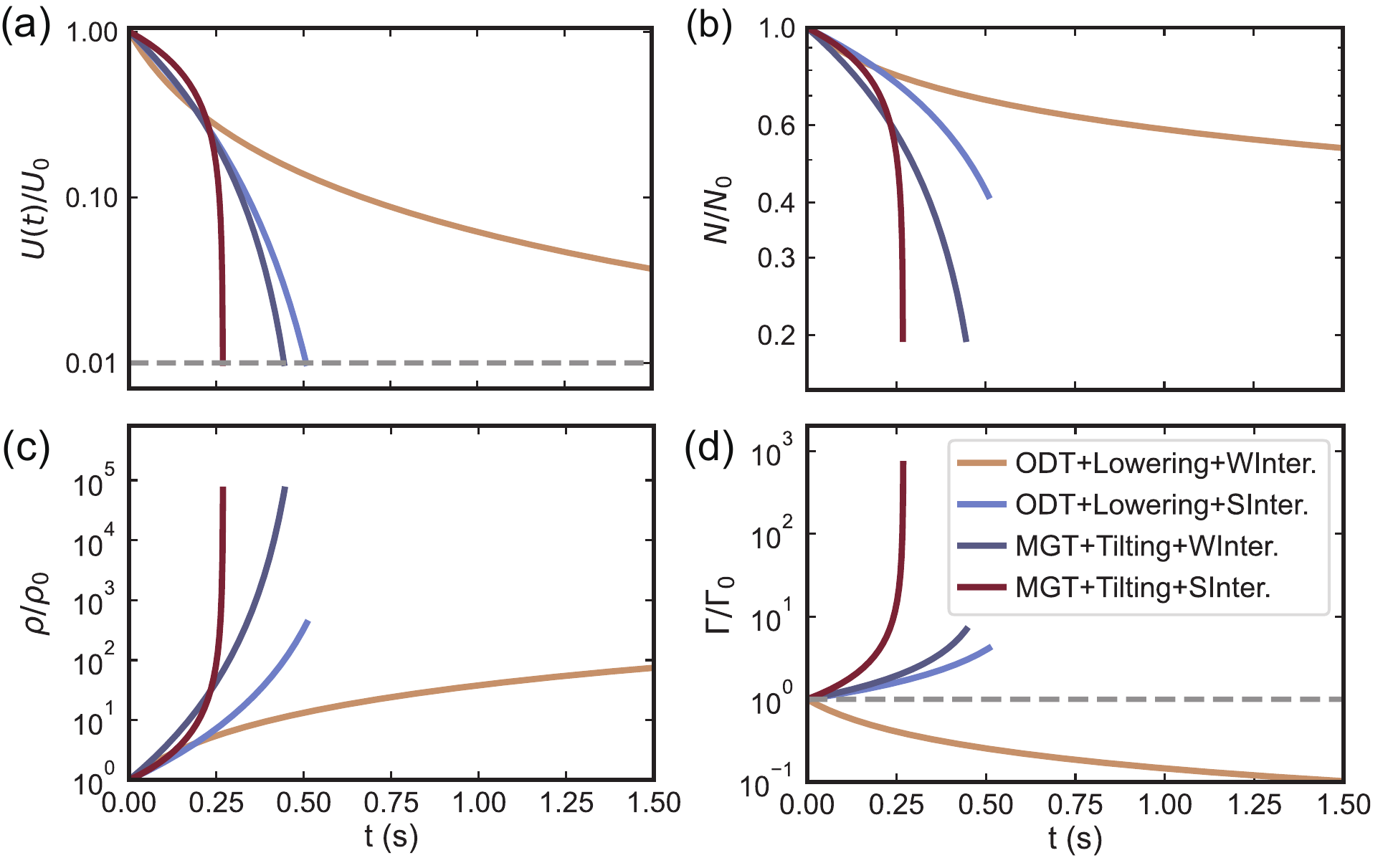}
		\caption{Simulated time evolution of $U(t)$, $N(t)$, $\rho(t)$, and $\varGamma(t)$ for the four evaporative-cooling schemes summarized in Table~\ref{tab:table_II}, using initial conditions taken prior to ODT lowering and a truncation parameter $\eta = 10$.
	} \label{fig:simulated_time_evo}
	\end{center}
\end{figure}

We define the evaporation rate to leading order in $e^{-\eta}$ as $\dot{N} \simeq -2(\eta - 4) e^{-\eta} \varGamma N$~\cite{O. J. Luiten1996}. 
By substituting the power-law scalings $N \propto U^q$ and $\varGamma \propto U^c$ derived in Appendix C into this rate equation, we obtain the time-dependent trap trajectory for constant $\eta$:
\begin{equation}\label{eq:tilt lowering curve}
	\frac{U(t)}{U_0} = \left( 1 - \frac{t}{\tau} \right)^{-1/p}.
\end{equation}
The exponent $p$ governs the curvature and is given by $p = 2 - 3\nu - q$. Note that for the unitary case, $p = -c$, directly linking the trajectory to the collision rate scaling. 
The characteristic time constant $\tau$ is determined by the initial collision rate $\Gamma_0$ and the trap geometry:
\begin{equation}
	\frac{1}{\tau} = -2(\eta - 4) e^{-\eta} \varGamma_0 \frac{p}{q}.
\end{equation}
With initial conditions taken before ODT lowering, Fig.~\ref{fig:simulated_time_evo} compares the dynamics for the schemes in Table~\ref{tab:table_II}.
The magnetically tilted evaporation ($\nu \approx 0$) exhibits the fastest trap reduction, reaching deep evaporation regimes within $\sim 0.25$~s. 
This rapid timescale arises from the favorable collision-rate scaling (small magnitude of $c$), which sustains a high $\varGamma$ even as $U$ drops. 
Consequently, this scheme yields the most rapid increase in phase-space density, significantly outperforming weakly interacting or purely optical lowering scenarios where the precipitous drop in collision rate limits the evaporation efficiency.

\renewcommand{\arraystretch}{1.6}
\begin{table*}
	\centering
	\caption{\label{tab:table_II}
		Trap-depth lowering curves $U(t)$ for different evaporative-cooling schemes, together with the associated characteristic timescale $1/\tau$ and initial collision rate $\varGamma_0$.}
	\begin{ruledtabular}
		\begin{tabular}{lccc}
			Scheme & $U(t)/U_0$ & $1/\tau$ & $\varGamma_0$ \\
			\hline
		    ODT+Lowering + WInter.~\cite{OHara2001} &
			$\left( 1 + \dfrac{t}{\tau} \right)^{-1/c}$ &
			$\dfrac{2}{3}\,\eta'\,(\eta - 4)\,e^{-\eta}\,\varGamma_0$ &
			$\dfrac{4\pi N_0 m \sigma \bar{\nu}_0^{3}}{k_\text{B} T_0}$ \\[2mm]
			ODT+Lowering + SInter~\cite{L. Luo2006} &
			$\left( 1 - \dfrac{t}{\tau} \right)^{1/c}$ &
			$\dfrac{2}{3}\,(\eta - 4)(\eta' - 6)\,e^{-\eta}\,\varGamma_0$ &
			$\dfrac{32\pi^2 \hbar^2 \bar{\nu}_0^{3} N_0 \eta}{U_{0}^{2}}$ \\[2mm]
			MGT+Tilting + WInter.~\cite{C.-L. Hung2008, J. F. Cl2009} &
			$\left( 1 + \dfrac{t}{\tau} \right)^{-1/c}$ &
			$\dfrac{2(\eta - 4)\,e^{-\eta}\,\varGamma_0\,c}{q}$ &
			$\dfrac{4\pi N_0 m \sigma \bar{\nu}_0^{3}}{k_\text{B} T_0}$ \\[2mm]
			\textbf{MGT+Tilting + SInter. (this work)} &
			$\mathbf{\left( 1 + \dfrac{t}{\tau} \right)^{-1/c}}$ &
			$\mathbf{\dfrac{2(\eta - 4)\,e^{-\eta}\,\varGamma_0\,c}{q}}$ &
			$\mathbf{\dfrac{32\pi^2 \hbar^2 \bar{\nu}_0^{3} N_0 \eta}{U_{0}^{2}}}$ \\
		\end{tabular}
	\end{ruledtabular}
\end{table*}

\section*{Appendix E: Expansion Dynamics and Thermometry}

To determine the temperature $T$ and the reduced temperature $T/T_F$ at the unitary limit, we reconstruct the initial in-trap mean square size $\langle r_i^2(0)\rangle$ from the time-of-flight (TOF) expanded size $\langle r_i^2(t)\rangle$ using the tilt-corrected trap frequencies $\omega_i$ derived in Appendix~A, which include the effects of the magnetic-gradient-induced tilt and residual magnetic curvature under the corresponding experimental conditions, via
\begin{equation}
	\langle r_i^2(0)\rangle = \frac{\langle r_i^2(t)\rangle}{b_i^2(t)},
\end{equation}
where $i\in\{x,y,z\}$. The evolution of the scaling factors $b_i(t)$ depends on the interaction regime.

For a strongly interacting gas, the expansion is hydrodynamic~\cite{J. E. Thomas2004}. The scaling factors are obtained by numerically integrating the hydrodynamic equations, which incorporate both $\omega_i$ and the residual magnetic curvature $\omega_{mi}$:
\begin{align}
	\ddot b_x + \omega_{mx}^2 b_x - \frac{\omega_x^2 + \omega_{mx}^2}{b_x \, \chi^{2/3}} &= 0 , \label{eq:hydro_x}\\
	\ddot b_y + \omega_{my}^2 b_y - \frac{\omega_y^2 + \omega_{my}^2}{b_y \, \chi^{2/3}} &= 0 , \\
	\ddot b_z - \omega_{mz}^2 b_z - \frac{\omega_z^2 - \omega_{mz}^2}{b_z \, \chi^{2/3}} &= 0 , \label{eq:hydro_z}
\end{align}
subject to initial conditions $b_i(0) = 1$ and $\dot b_i(0)=0$. Here, $\chi \equiv b_x b_y b_z$ is the volume scaling factor coupling the three axes. 

For noninteracting clouds, the expansion is ballistic. The corresponding scaling factors are obtained from Eqs.~(\ref{eq:hydro_x}--\ref{eq:hydro_z}) by substituting the coupling term $\chi$ with $b_i^3$ for each axis $i$, which decouples the equations and recovers the standard ballistic scaling ($\propto 1/b_i^3$).

Using the virial theorem, the temperature of a unitary Fermi gas is related to the in-trap size by~\cite{thomasVirial2008}
\begin{equation}
	k_B T = m \omega_i^2 \langle r_i^2(0)\rangle,
\end{equation}
Note that formally $T$ for a unitary gas is defined via an effective mass or frequency, but algebra leads to the simple relation above using the physical mass $m$.
However, the Fermi temperature $T_F$ is rescaled due to many-body interactions. The reduced temperature is given by
\begin{equation}
	\frac{T}{T_F} = \frac{m \omega_i^2 \langle r_i^2(0)\rangle}{\hbar \sqrt{1+\beta}\, \bar\omega (6N)^{1/3}},
\end{equation}
where $\beta$ is a universal many-body parameter. For a unitary gas, we use $\beta = -0.49$~\cite{J. Kinast2005}.  

\newpage

\section*{Appendix F: Tilted Cooling of a Weakly Interacting Fermi Gas}

To demonstrate the generality of the protocol, we apply magnetic-tilt evaporation to a weakly interacting gas at a bias field of 320\,G. 
Following a 2.7\,s exponential ramp-down of the ODT, we initiate the tilt sequence while maintaining constant laser intensity. 

\begin{figure}[h]
	\begin{center}
		\includegraphics[width=0.95\columnwidth, angle=0]{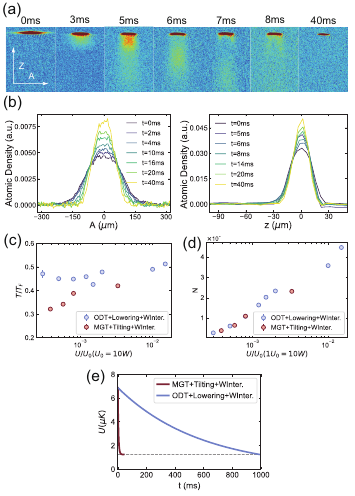}
		\caption{Experiment and results of the tilt-cooled weakly interacting Fermi Gas. 
		The experimental conditions are the same as in Fig.~\ref{fig:experimental_procedure}(a), except that the gas is operated in the weakly interacting regime. (c) and (d) present the obtained $T/T_F$ and $N$, respectively.
		(e) The ODT-lowering curve is adjusted accordingly to match the appropriate scaling. 
		} \label{fig:tilt_cooling_weak_int_time}
	\end{center}
\end{figure}

The advantage of this method is evident when comparing the cooling limits, as shown in Fig.~\ref{fig:tilt_cooling_weak_int_time}. 
In conventional optical evaporation, lowering the trap depth to $U = 0.6\,\mu\mathrm{K}$ reduces the temperature, yet the degeneracy saturates at $T/T_F \approx 0.42$. 
In contrast, using the tilted protocol starting from a deeper ODT ($U = 1.2\,\mu\mathrm{K}$) and applying a gradient of 0.32\,$g$ to reach an effective depth of $U_{\text{eff}} \approx 0.6\,\mu\mathrm{K}$, we achieve a significantly lower reduced temperature of $T/T_F \approx 0.32$. 
At this point, while the atom number and absolute temperature are comparable to the purely optical case at $0.7\,\mu\mathrm{K}$, the reduced temperature is markedly lower. 
These results confirm that magnetic tilting bypasses the stagnation of cooling efficiency encountered in shallow optical traps, providing access to deeper degeneracy even in the weakly interacting regime.
\newpage

\end{document}